\documentclass[11pt,a4paper]{article}
\usepackage{jheppub}

\usepackage{graphicx}
\usepackage{amsmath}
\usepackage{amssymb}
\usepackage{braket}
\usepackage{url}
\usepackage{comment}
\usepackage{float}
\usepackage{color}

\usepackage[utf8]{inputenc}

\usepackage{subfigure}
\newcommand{\bea}{\begin{eqnarray}}
\newcommand{\halpha}{\hat{\alpha}}

\newcommand{\eea}{\end{eqnarray}}
\newcommand{\be}{\begin{equation}}
\newcommand{\ee}{\end{equation}}
\newcommand{\ba}{\begin{align}}
\newcommand{\ea}{\end{align}}
\newcommand{\V}{\mathcal{V}}

\newcommand{\Kahler}{\ensuremath{\text{K}\ddot{\text{a}}\text{hler}\,}}

\def\pref#1{(\ref{#1})}
\newcommand{\ma}{m_{\scriptscriptstyle A}}
\newcommand{\mb}{m_{\scriptscriptstyle B}}
\def\bel#1{\begin{equation} \label{#1}}

\title{Hidden Sectors in String Theory:\\ Kinetic Mixings, Fifth Forces and Quintessence}

\author[1,2]{Bobby Samir Acharya}
\author[3]{, Anshuman Maharana}
\author[2]{and Francesco Muia}

\affiliation[1]{\small \it Theoretical Particle Physics \& Cosmology Group, Department of Physics, King’s College London, Strand, London, WC2R 2LS, United Kingdom}
\affiliation[2]{\small \it ICTP, Strada Costiera 11, Trieste 34014, Italy}
\affiliation[3]{\small \it Harish Chandra Research Institute,
Homi Bhabha National Institute, Jhunsi, Allahabad, Uttar Pardesh, India 211019}

\abstract{
Light moduli fields in string compactifications can have interesting implications for particle physics and cosmology. Fifth force bounds impose stringent constraints on the interactions of such moduli with the visible sector. To be consistent with the bounds, they need to  be part of hidden sectors which interact with the Standard Model with weaker-than-Planck suppressed interactions. We consider scenarios in which the visible sector degrees of freedom are localised in the compactification and light moduli arise as closed string degrees of freedom associated with hidden sectors which are geometrically separated (in the extra-dimensions) from the Standard Model. Kinetic mixings lead to interactions between the moduli and the visible sector - we compute these using \Kahler potentials of string/M-theory compactifications. We argue that in general these interactions provide a lower bound on the strength of the interactions between the moduli and the visible sector. The interactions scale with inverse powers of the volume of the compactification, thus fifth force bounds can be translated to lower bounds on the volume of the extra-dimensions. We find that compactification volumes have to be large to evade the bounds. This imposes interesting constraints on quintessence model building in string theory. Our results for the strength of the interactions can also be used to quantify the fine-tuning necessary for the stability of the potential of a light modulus against quantum corrections involving visible sector loops.
}

\keywords{String Theory, Moduli, Fifth Forces,  Quintessence, Swampland.}

\arxivnumber{}

\date{\small\today}

\begin{document} 
\maketitle

\sloppy
\section{Introduction}

\label{sec:intro}

Moduli fields are a generic feature of string compactifications.  Typically, they couple to the visible sector via Planck suppressed interactions; fifth force bounds then require their masses to be above the meV scale (see e.g.  \cite{Will:2014kxa, Adelberger:2003zx}). This bound is usually not considered as a challenge for string phenomenology since effective field theory arguments tie moduli masses to the scale of supersymmetry breaking. Thus, even if supersymmetry is responsible for stabilising the weak scale, the lightest moduli are expected to have masses of the order of a TeV. Although they do not mediate any long-range forces, such moduli have a significant impact on cosmology (as vacuum misalignment leads to an epoch in which the energy density of the universe is dominated by cold moduli particles). This leads to interesting phenomenological scenarios which have correlations  between  scales in particle physics and cosmology \cite{mod1, ModuliCosmology}. 

One should keep in mind that the above line of reasoning relies on effective field theory arguments and naturalness considerations. The extreme smallness of the observed value the cosmological constant has put the doctrine of naturalness under question. Given this, it is worth  exploring the phenomenology of string compactifications with light moduli that can mediate long-range forces. For such theories, fifth force bounds impose very stringent constraints \cite{Will:2014kxa, Adelberger:2003zx, Jain:2010ka, Damour:2010rp, Touboul:2017grn, Berge:2017ovy, Wagner:2012ui, Nobili:2018eym}. The couplings between the moduli and the visible sector have to be significantly
weaker than that of gravitational interactions. Thus, the moduli have to arise from hidden sectors which interact very weakly with the Standard Model. Hidden sectors are generic in string theory and are often necessary for the consistency of compactifications, see for example \cite{Giedt:2000bi} for a discussion in the context of heterotic orbifold models, \cite{Cvetic:2004ui} in type II, \cite{Taylor:2015ppa} in F-theory and \cite{Acharya:2016fge, Acharya:2017kfi} in M-theory. Moduli associated with such sectors can have interesting dynamics which can be important for late time cosmology.


There has been steady progress in our understanding of moduli dynamics and their cosmological implications. In this context, the simplest possibility is to  stabilise  all the moduli at a de Sitter minimum. The most well-developed models for such a scenario are in type IIB \cite{Kachru:2003aw, Denef:2004dm, Denef:2005mm, Lust:2005dy, Lust:2006zg, Balasubramanian:2005zx, Cicoli:2013cha, Westphal:2006tn, Louis:2012nb, Rummel:2011cd, Braun:2015pza, Gallego:2017dvd}, where fluxes threading the internal cycles are responsible for a large multitude of solutions \cite{Dasgupta:1999ss, Gukov:1999ya, Bousso:2000xa, Feng:2000if, Giddings:2001yu}. Models in M-theory  were constructed in \cite{Acharya:2006ia}.  For constructions in heterotic strings see \cite{het}, and for constructions in type IIA see \cite{typeIIa}. Constructions in  non-critical strings  have been carried out in \cite{Maloney:2002rr}. There have also been efforts to construct models of quintessence in string/M-theory \cite{Choi:1999xn, Kaloper:2008qs, Panda:2010uq, Cicoli:2012tz, Olguin-Tejo:2018pfq}. Of these, the ones with moduli stabilisation are closely related to the de Sitter constructions and make use of the same approximations. 

At the same time, a conjecture has been put forward  which puts all de Sitter vacua in the swampland \cite{Obied:2018sgi} (for earlier criticisms of dS constructions see e.g. \cite{Banks:2012hx, Bena:2009xk, Bena:2012bk, Moritz:2017xto, Sethi:2017phn, Danielsson:2018ztv}) and implies that the current cosmic acceleration is driven by quintessence. So far, the evidence presented in favour of the conjecture has been only in classical string  theory \cite{Obied:2018sgi} and regimes of parametric weak coupling \cite{Ooguri:2018wrx}, see \cite{Cicoli:2018kdo, Kachru:2018aqn, Kallosh:2018wme, Akrami:2018ylq} for recent reassessments of de Sitter constructions and critical discussions of the conjecture. Related explorations have been carried out in \cite{Agrawal:2018own, Dvali:2018fqu, Andriot:2018wzk, Banerjee:2018qey, Achucarro:2018vey, Garg:2018reu, Kehagias:2018uem, Dias:2018ngv, Denef:2018etk, Colgain:2018wgk, Roupec:2018mbn, Andriot:2018ept, Matsui:2018bsy, Ben-Dayan:2018mhe, Damian:2018tlf, Conlon:2018eyr, Kinney:2018nny, Dasgupta:2018rtp, Murayama:2018lie, Brahma:2018hrd, Choi:2018rze, Das:2018hqy, Danielsson:2018qpa, Wang:2018duq, Han:2018yrk, Moritz:2018ani, Bena:2018fqc, Dimopoulos:2018upl, Heisenberg:2018rdu, Heisenberg:2018yae, DAmico:2018mnx, Ashoorioon:2018sqb, Odintsov:2018zai, Motaharfar:2018zyb, Kawasaki:2018daf, Hamaguchi:2018vtv, Lin:2018kjm, Ellis:2018xdr, Das:2018rpg, Wang:2018kly, Fukuda:2018haz, Hebecker:2018vxz, Gautason:2018gln,  Garg:2018zdg, Park:2018fuj, Blaback:2018hdo, Schimmrigk:2018gch, Lin:2018rnx, Agrawal:2018rcg, Yi:2018dhl, Heckman:2018mxl, Chiang:2018lqx, Cheong:2018udx, Elizalde:2018dvw, Blanco-Pillado:2018xyn, Ibe:2018ffn, Holman:2018inr, Junghans:2018gdb, Banlaki:2018ayh, Andriot:2018mav, Wang:2018fng, Capozziello:2018jya, Dutta:2018vmq, Russo:2018akp}.


If the present-day dark energy is to be attributed to quintessence (see \cite{Tsujikawa:2013fta} for a recent review of quintessence), then the associated scalar has Compton wavelength of the order of the cosmological horizon. The field is effectively massless and the stringent fifth force bounds described above need to be addressed\footnote{In addition, bounds on the time variation of fundamental constants impose further constraints \cite{Martins:2017yxk}. From a  theoretical perspective, stabilising the potential of the quintessence field against quantum corrections requires fine-tuning at the functional level \cite{Banks:2001qc, Garny:2006wc, Marsh:2018kub}.} (see e.g.  \cite{Carroll:1998zi}). As a possible resolution to the problem, it has been proposed that the quintessence field can be a modulus field  in a hidden sector which is geometrically separated from the Standard Model sector \cite{Agrawal:2018own, Obied:2018sgi}. Geometric separation implies that the wavefunctions of the modes in the extra-dimensions have negligible overlap and hence leads to the absence of certain couplings in the tree level (super)potential. But the sequestering is never complete - higher derivative corrections, loop effects and kinetic mixings invariably lead to interactions between sectors that are geometrically separated \cite{mixkac}. While small, such interactions can have important phenomenological implications: in some situations  they can be used to generate small numbers, while in others they can generate matrix elements that can be dangerous for phenomenology.

The goal of this paper is to analyse the strength of interactions between the Standard Model sector and geometrically separated moduli fields. We will examine the mixings in the context of fifth force bounds, and  translate the bounds to criteria on the geometry of  compactifications. This will give us lower bounds on the volume of compactifications. We analyse mixings that arise between scalars and mixings of $U(1)$ gauge fields. In the case of the scalars, the mixings arise from the diagonalisation of the kinetic and mass matrices. We will argue that in general the mixing from diagonalising the kinetic term provides a lower bound on the strength of the interactions (unless the kinetic and mass matrices are aligned). Given this, in order to keep our results as model-independent as possible we will not commit to any potential for the scalars - we will consider the known form of the \Kahler potential in various string/M-theory constructions and obtain the lower bounds they imply on the strength of interactions between a geometrically separated modulus and the Standard Model sector. The \Kahler potentials we will use are valid in the limit of weak coupling and large volume, this is also the regime in which the effects of geometric separation are expected to be maximal. We find that even the lower bounds imposed by them give interesting constraints on  model building\footnote{Our results should also be interesting in the context of the scalar version of the weak gravity conjecture \cite{Palti:2017elp}, although we do not explore this in detail.}. Our analysis also reveals that in some constructions the geometrically separated modulus couples to different degrees of freedom of the Standard Model
with different strengths. Detailed analysis of the implications of a scalar mediated  fifth force for violations of the equivalence principle has been only carried out for the cases in which the scalar couples with a universal strength to the Standard Model degrees of freedom. Our results motivate a comprehensive study of the phenomenology when the couplings are non-universal.

This paper is organised as follows. Section \ref{sec:fifth} reviews fifth forces mediated by light scalars: the basic formalism used for their study and the experimental bounds. Section \ref{sec:kin} discusses the couplings that arise as a result of kinetic mixing of scalar moduli. Here, we begin with a general discussion which illustrates how potentially dangerous couplings can arise from kinetic mixing. We then go on to  specific examples and obtain the strengths of the couplings that are induced. The computations will use the same methods as in \cite{Conlon:2007gk}, although there the analysis was carried out in the presence of a specific potential - both the kinetic and mass matrices were diagonalised (\cite{Cicoli:2010ha} also diagonalises both kinetic and mass matrices in a large class of examples). Having obtained the strength of the interactions we will use the observational bounds on fifth forces mediated by light scalars to discuss the implications of our results for model building (particularly in the context of quintessence). Section \ref{sec:uone} deals with kinetic mixing of gauge bosons. Section \ref{sec:disc} gives a general discussion of our results and future directions. 

\section{Light Scalars and Fifth Forces} 

\label{sec:fifth}

Fifth forces are ubiquitous in Beyond the Standard Model (BSM) theories. For detailed reviews of the experimental efforts to detect fifth forces, the bounds and the theoretical origin of fifth forces in various BSM scenarios see e.g. \cite{Will:2014kxa, Adelberger:2003zx, Jain:2010ka}. Our discussion shall be in the context of light scalars, in particular the fifth force that would be mediated by a scalar field driving quintessence. Recall that the mass of  the field driving quintessence has to be of the order of $10^{-32} \phantom{a} \text{eV}$ and the corresponding Compton wavelength is approximately $10^{25} \phantom{s} \text{m}$. This is almost of the size of the observable universe: for the purposes of studying the effects that such a scalar can have on the violations of the equivalence principle it can be taken to be massless. The basic formalism for analysing the violations of the equivalence principle that can be induced by a scalar (both massive and massless) was developed in \cite{Damour:2010rp} (for a qualitative and simplified discussion in the context of quintessence we refer the reader to \cite{Carroll:1998zi}). Here we briefly review the results of \cite{Damour:2010rp} that shall be relevant for our discussion.

The starting point for the analysis of \cite{Damour:2010rp} is the effective action for the Standard Model and the scalar at energy scales
slightly above $1 \ {\rm{GeV}}$. The relevant degrees of freedom in the Standard Model are the up $(u)$ and down $(d)$ quarks, the electron $(e)$, the photon $A_{\mu}$ and the gluonic gauge fields $A_{\mu}^{A}$ (it can be argued that the effects related to the strange quark are negligible). The couplings of the scalar $(\chi)$ to the Standard Model degrees of freedom are characterised by five parameters $d_i$ $(d_e, d_g, d_{m_{e}}, d_{m_{u}}$ and $d_{m_{d}})$ which appear in the interactions of the scalar:
 \begin{equation}
 \label{phiint}
   \mathcal{L}_{\rm{int}\chi} = {\chi \over M_{\rm pl} }\left[ +{d_e  \over 4 e^2 } F_{\mu \nu} F^{\mu \nu} - { {d_g \beta_3(g_3)} \over {2 g_3} } F_{\mu \nu}^{A} F^{A \mu \nu}
  - \sum_{e,u,d} \left(d_{m_i} + \gamma_{m_{i}} d_g \right) m_i \bar{\psi}_i \psi_i \right] \,,
 \end{equation}
 where $\beta_3(g_3) = \mu \partial g_3(\mu)/ \partial \mu$ is the QCD $\beta$-function that governs the running of $g_3$ and the second term in Eq. \eqref{phiint} is given by the QCD trace anomaly.
Reference \cite{Damour:2010rp} analysed violations of the equivalence principle that can arise as a result of the above interactions and found that the violations induced are a function of the mass of the scalar, composition of the test bodies and a four dimensional subspace of the five dimensional parameter space.
 
The interactions in Eq. \pref{phiint} are defined at a low scale $(\mu \sim 1 \phantom{a} \rm{GeV})$. They are supposed to be
determined from RG evolution of a high scale Lagrangian derived from string (or any other UV complete) theory, defined at a cut-off scale $\Lambda_c$. The UV Lagrangian would contain the terms:
\begin{equation}
\label{UVLag}
   \mathcal{L}_{\Lambda_c} \supset   -  {1 \over  {4 e^2(\Lambda_c, \chi)}} F_{\mu \nu} F^{\mu \nu} - {1 \over  {4 g_3^2(\Lambda_c, \chi)}}  F_{\mu \nu}^{A} F^{A, \mu \nu}
   - \sum_{e,u,d}  m_i (\Lambda_c, \chi) \bar{\psi}_i \psi_i \,.   
\end{equation}
The UV interaction strengths can be defined by introducing the parameters $(d^c_{i})$:
\be
\label{UVint}
  d^{c}_e  = M_{\rm pl} { {\partial \ln e^2(\Lambda_c, \chi)} \over \partial{\chi}},  \qquad
  d^{c}_g  = M_{\rm pl} { {\partial \ln g_3^2(\Lambda_c, \chi)} \over \partial{\chi}},  \qquad
  d^{c}_{m_{i}}  = M_{\rm pl} { {\partial \ln m_{i}(\Lambda_c, \chi)} \over \partial{\chi}}.
\ee
While the precise relationships between the UV interaction strengths $(d^{c}_i)$ and the low energy parameters $(d_i)$ will depend on the details of the theory, \cite{Damour:2010rp} argued on general  grounds that
\be
\label{rela}
   d_e \sim d_e^c, \qquad d_g \sim K d_g^c , \qquad  d_{m_{i}} \sim K_{m_{i}} d^c_{m_{i}} \,,
 \ee
where $K, K_{m_{i}}$ are constants of the order of $40$. A more detailed analysis of violations of the equivalence principle was carried out assuming that all the UV interactions are of the same magnitude 
(as is true in many string theory examples), i.e.
$$
  d_e^{c} \sim d_g^c \sim d^c_{m_{i}} \equiv d^c,
$$
In this case, it was found that the equivalence principle violating effects can be parametrised in terms of only two variables (which are functions of the $d_i$ and the atomic weights and numbers of the test bodies) and the mass of the scalar. Using the results of the E\"{o}tWash experiment \cite{Schlamminger:2007ht} and celestial Lunar Laser Ranging \cite{Williams:2004qba}, for massless scalars ref. \cite{Damour:2010rp} obtained the bound 
\begin{equation}
\label{boundt}
 \left(d^c\right)^2 < 10^{-12}.
\end{equation}
More recently, data from the MICROSCOPE mission \cite{Touboul:2017grn} has been analysed using the two variable parametrisation of \cite{Damour:2010rp}. Consequently, the above bound has become stronger by one order of magnitude \cite{Berge:2017ovy}. 
 
\section{Kinetic mixing of Scalars}

\label{sec:kin}

\subsection{General argument}

In this subsection, we  argue that in general geometric separation does not lead to complete sequestering between a modulus and the visible sector, as kinetic terms always produce mixings. Consider $N$ scalar fields $\phi^I$ with the Lagrangian
\begin{equation}
\mathcal{L} = K_{IJ} (\phi^{I}) \partial_\mu \phi^I \partial^\mu \phi^J - V(\phi_I) \,,
\end{equation}
where the kinetic matrix $K_{IJ}$ is positive definite and $I, J = 1, \dots, N$. In string/M-theory compactifications, the kinetic and  potential terms are derived from the \Kahler and superpotential of the construction. These are computed in the geometric basis for the scalars, that leads to the absence of direct couplings between geometrically separated sectors in the tree level superpotential. In order to obtain physical couplings, canonical normalisation has to be carried out. At any specific point in field space one can write the fields
 $\phi^{I}$ as sums of their expectation values and fluctuations: $\phi^{I} = \langle \phi^I \rangle + \delta \phi^{I}$, then carry out a linear change in basis which takes the fluctuations in the geometric basis ($\delta \phi^{I}$) to the canonical ones ($\varphi^{I}$):
\begin{equation}
\delta \phi^I = \mathcal{M}^I_J \varphi^J \,,
\end{equation}
where the matrix $\mathcal{M}^{I}_{J}$ satisfies
\begin{equation}
\label{bchange}
K_{IJ} \mathcal{M}^I_K \mathcal{M}^J_L = \delta_{KL} \,,
\end{equation}
so that the kinetic terms becomes diagonal. We note that the matrix $\mathcal{M}^{I}_{J}$ is easily obtained from the eigenvectors of kinetic matrix. The condition in Eq. \pref{bchange} can be satisfied by taking
\begin{equation}
\label{eq:RotationGeneral}
\mathcal{M}^I_J = \frac{e^I_J}{\sqrt{\lambda_J}} \,,
\end{equation}
where $e^{I}_{J}$ is the $J^{\rm th}$ eigenvector (normalised to unity) of the kinetic matrix and $\lambda_J$ is the corresponding eigenvalue.

Consider a situation where in the geometric basis a certain Standard Model coupling is determined by particular field a ($\phi^A$), for instance the gauge coupling of D7-branes wrapping the cycle $A$:
\begin{equation}
\label{eq:CouplingGaugeQ}
{ \mathcal{L} } \supset  \phi^{\large{A}} F^{a}_{\mu \nu} F^{a, \mu \nu} \ ,
\end{equation}
On making the basis change to the canonical basis this leads to a term in the Lagrangian:
\begin{equation}
{ \mathcal{L} } \supset  M^{\large{A}}_{\phantom{A}J} \varphi^{J} F^{a}_{\mu \nu} F^{a, \mu \nu}.
\end{equation}
Note that in the new basis, the strength of the coupling of a (geometrically separated) scalar $\varphi^{B}$ to the gauge fields is determined by the magnitude of the off-diagonal entry $M^{\large{A}}_{\phantom{A}B}$. Thus, potentially dangerous couplings
between the Standard Model gauge bosons and a geometrically separated scalar can be generated. Similarly, the Standard Model Yukawas can also acquire dependence on hidden sector scalars. We will examine both dependences in detail in the examples below.

After canonically normalising kinetic terms, the  mass matrix has to be diagonalised. This basis change depends on the potential for the scalars. If the mixings induced by this are of smaller magnitude than those induced by the basis change required for canonical normalisation of the kinetic terms, they can be neglected. On the other hand, if the mixings that arise from diagonalising the mass matrix are of larger magnitude, then the interactions induced are of greater strength than those obtained from  canonical normalisation of the kinetic terms. If both basis change matrices have off-diagonal entries of the same order of magnitude, the strength of interactions is in general of the order of magnitude given by those obtained from canonical normalisation of the kinetic matrix. Thus, unless the kinetic and mass matrices are aligned so that their effects precisely cancel, the strength of the mixings
after diagonalising the mass matrix can only \textit{increase}. Therefore, in order to make our study model independent we will not commit to any potential for the scalars. We will consider \Kahler potentials in various compactifications and obtain the strengths of the mixings they induce\footnote{See \cite{Conlon:2007gk, Cicoli:2010ha} for computations where both kinetic and mass matrices are diagonalised. The basis mechanism behind the mixings is similar.}. Our results should therefore be considered as {\it lower bounds} on the strengths of the interactions.

Given the constraints from fifth force bounds and time variation of fundamental constants, our results have interesting implications for model building with light scalars (particularly in the context of quintessence). The above arguments imply that there are essentially two ways to avoid dangerous couplings between the Standard Model sector and a light scalar: 
\begin{itemize}
\item The size of the off-diagonal entries in the basis change matrix (which induce the coupling between the scalar and the visible sector) are small. This will impose lower bounds on the volume of the compactification in the examples that we consider below. 
\item There is negligible coupling between the light direction in field space and the Standard Model. This requires tuning. Firstly, this would require an alignment between the kinetic and mass matrices (as described above). Secondly, in general different Standard Model degrees of freedom couple to different directions in the scalar field space (as we will see in the examples below). Thus the absence of couplings to all the degrees of freedom would require further tuning.  
     \end{itemize}
We now turn to the analysis of kinetic mixing in specific settings. We shall consider examples where the geometrically separated scalar is a blow-up mode or fibre modulus. The visible sector will be realised by branes wrapping a blow-up cycle or from branes at singularities. It will suffice to consider semi-realistic models of the visible sector for our purposes.

\subsection{Blow-up Models}

Blow-up moduli, corresponding to resolutions of point-like singularities, have their wavefunctions localised in the internal manifold. If the visible sector degrees of freedom are localised away from the resolution, then it is natural to expect that they will interact with the blow-up of a point-like singularity weakly. In this subsection, we will take the light scalar (candidate to be driving quintessence) to be a blow-up mode. We start by looking at IIB  string theory, where we consider two examples: the case when the Standard Model is realised from D7-branes wrapping another blow-up mode (in the geometric regime) and the case when it is realised from branes at singularities.  We then study an example in M-theory with a single blow-up mode.

\subsubsection{SM at a geometric blow-up}
\label{twoblow}

In type IIB Calabi-Yau compatifications, the \Kahler potential is given by
\begin{equation}
    K = -2  \log {\mathcal{V}},
\end{equation}
where ${\mathcal{V}}$ is the volume of the compactification. Consider  a Swiss-cheese type Calabi-Yau which has three \Kahler moduli: with $\tau_0$ as the big cycle and $\tau_{1,2}$ as two blow-ups.  We  will work with a setup where the Standard Model degrees of freedom will be localised on $\tau_1$ and the role of geometrically separated light modulus is played by $\tau_2$. The \Kahler potential for the moduli takes the form:
\begin{equation}
K = -2 \log\left(\alpha\left(\tau_0^{3/2} - \gamma_1 \tau_1^{3/2} - \gamma_2 \tau_2^{3/2}\right)\right) \,,
\end{equation}
where  $\alpha$, $\gamma_1$ and $\gamma_2$ are constants\footnote{For explicit realisations in weighted projective spaces see  e.g. \cite{Blumenhagen:2007sm}.}. In the large volume limit, $\V \simeq \tau_0^{3/2} \gg 1$, we can work perturbatively in $\epsilon \equiv \tau_0^{-1} \ll 1$. The \Kahler metric can be expanded as
\begin{equation}
\label{eq:DecomposedMetric}
K_{ij} = K^{(0)}_{ij} + K^{(1)}_{ij} + K^{(2)}_{ij} + . . . \,,
\end{equation}
with
\begin{equation}
K^{(0)}_{ij} = 
\begin{pmatrix}
A \epsilon^2 && 0 && 0 \\
0 && B \epsilon^{3/2}  && 0 \\
0 && 0 && C \epsilon^{3/2}
\end{pmatrix} \,, \nonumber
\end{equation}
\begin{equation}
K^{(1)}_{ij} = 
\begin{pmatrix}
0 && D \epsilon^{5/2} && E \epsilon^{5/2}  \\
D \epsilon^{5/2} && 0 && 0 \\
E \epsilon^{5/2} && 0 && 0
\end{pmatrix} \,, \quad
K^{(2)}_{ij} =
\begin{pmatrix}
0 && 0 && 0 \\
0 && 0 && F \epsilon^{3} \\
0 && F \epsilon^{3} && 0 ,
\end{pmatrix} \ ,
\end{equation}
where  we have defined
\begin{equation}
A = \frac{3}{4} \,, \qquad B = \frac{3 \gamma_1}{8 \sqrt{\tau_1}} \,, \qquad C = \frac{3 \gamma_2}{8 \sqrt{\tau_2}} \,, \nonumber
\end{equation}
\begin{equation}
D = -\frac{9}{8} \gamma_1 \sqrt{\tau_1} \,, \qquad E = -\frac{9}{8} \gamma_2 \sqrt{\tau_2} \,, \qquad F = \frac{9}{8} \gamma_1 \gamma_2 \sqrt{\tau_1 \tau_2} \,.
\end{equation}
The unperturbed eigenvalues (denoted by superscript $(0)$) can be read then from the diagonal entries of the matrix $K^{(0)}$ and are
\begin{equation}
\lambda_0^{(0)} = A \epsilon^2 \,, \qquad \lambda_1^{(0)} = B \epsilon^{3/2} \,, \qquad \lambda_2^{(0)} = C \epsilon^{3/2} \,,
\end{equation}
corresponding to the unperturbed eigenvectors
\begin{equation}
\mathcal{B}^{(0)} = \left\{\begin{pmatrix} 1 \\ 0 \\ 0 \end{pmatrix}, \begin{pmatrix} 0 \\ 1 \\ 0 \end{pmatrix}, \begin{pmatrix} 0 \\ 0 \\ 1 \end{pmatrix}\right\} \,.
\end{equation}
Recall that non-degenerate perturbation theory is good as long as the splittings in the unperturbed eigenvalues are larger than the size of the perturbations. We will assume that we are away from special points in moduli space where the splittings are small or comparable to the perturbations, and use non-degenerate perturbation theory to compute the eigenvalues and eigenvectors of $K_{ij}$\footnote{For special points in the moduli space where the splittings are small in comparison with the perturbations, our results can be easily generalised using degenerate perturbation theory. Here we note that typically break down of non-degenerate perturbation theory implies that the perturbation leads to stronger mixings between the unperturbed eigenvectors.}. 

Consider the first perturbation, $ K_{ij}^{(1)}$: the perturbed eigenvalues can be computed in perturbation theory by solving the equation
\begin{equation}
\det\left(K_{ij}^{(0)} + \epsilon K_{ij}^{(1)} - \lambda \delta_{ij}\right) = 0 \,.
\end{equation}
As expected for off-diagonal corrections connecting non-degenerate eigenvalues, the correction appears at $\mathcal{O}(\epsilon^2)$ in perturbation theory. We define
\begin{align}
\lambda^{(1)}_0 = \lambda_0^{(0)} + \delta_1 \lambda_0 \,, \qquad \lambda^{(1)}_1 = \lambda_1^{(0)} + \delta_1 \lambda_1 \,, \qquad \lambda^{(1)}_2 & = \lambda_2^{(0)} + \delta_1 \lambda_2 \,.
\end{align}
Then the corrections are
\begin{equation}
\delta_1 \lambda_0 \simeq \frac{B E^2 + C D^2}{BC} \epsilon^{7/2} \,, \qquad \delta_1 \lambda_1 \simeq \frac{D^2}{B} \epsilon^{7/2} \,, \qquad \delta_1 \lambda_2 \simeq \frac{E^2}{C} \epsilon^{7/2} \,.
\end{equation}
Using these results we can find the first-order perturbed eigenvectors:
\begin{equation}
\mathcal{B}^{(1)} = \left\{ v_0^{(1)} = \begin{pmatrix} 1 \\ \beta_0^{(1)} \\ \gamma_0^{(1)} \end{pmatrix}, v_1^{(1)} =  \begin{pmatrix} \alpha_1^{(1)} \\ 1 \\ \gamma_1^{(1)} \end{pmatrix}, v_2^{(1)} =  \begin{pmatrix} \alpha_2^{(1)} \\ \beta_2^{(1)} \\ 1 \end{pmatrix} \right\} \,,
\end{equation}
where
\begin{equation}
\beta_0^{(1)} = -\frac{D}{B} \epsilon \,, \qquad \gamma_0^{(1)} = - \frac{E}{C} \epsilon \,, \nonumber
\end{equation}
\begin{equation}
\alpha_1^{(1)} = \frac{D}{B} \epsilon \,, \qquad \gamma_1^{(1)} = \frac{ED}{B(B-C)} \epsilon^2 \,, \nonumber
\end{equation}
\begin{equation}
\label{eq:Eigenvec15}
\alpha_2^{(1)} = \frac{E}{C} \epsilon \,, \qquad \beta_2^{(1)} = \frac{ED}{C(C-B)} \epsilon^2 \,.
\end{equation}
Next, we compute the corrections induced by $K^{(2)}_{ij}$. They can be computed by simply requiring that
\begin{equation}
\label{eq:EigenvaluesEquation32}
\left(K_{ij}^{(0)} + K_{ij}^{(1)} + K_{ij}^{(2)}\right) v^{(2)}_j = \lambda^{(1)}_i v^{(2)}_i\,,
\end{equation}
where
\begin{equation}
v^{(2)}_i = v_i^{(1)} + \delta_2 v_i \,, \quad i=1,2,3 \,,
\end{equation}
and
\begin{equation}
v_0^{(2)} = \begin{pmatrix} 1 \\ \beta_0^{(1)} \\ \gamma_0^{(1)} \end{pmatrix} + \begin{pmatrix} \delta_2 \alpha_0 \\ \delta_2 \beta_0 \\ \delta_2 \gamma_0 \end{pmatrix} \,, \quad v_1^{(2)} = \begin{pmatrix} \alpha_1^{(1)} \\ 1 \\ \gamma_1^{(1)} \end{pmatrix} + \begin{pmatrix} \delta_2 \alpha_1 \\ \delta_2 \beta_1 \\ \delta_2 \gamma_1 \end{pmatrix} \,, \quad v_2^{(2)} = \begin{pmatrix} \alpha_2^{(1)} \\ \beta_2^{(1)} \\ 1 \end{pmatrix} + \begin{pmatrix} \delta_2 \alpha_2 \\ \delta_2 \beta_2 \\ \delta_2 \gamma_2 \end{pmatrix} \,,
\end{equation}
The only non-subleading contributions are
\begin{equation}
\delta_2 \gamma_1 = \frac{F}{B-C} \epsilon^{3/2} \,, \qquad \delta_2 \beta_2 = \frac{F}{C-B} \epsilon^{3/2} \,.
\end{equation}

Having obtained the eigenvalues and eigenvectors, let us compute the basis change which relates the geometrical moduli to the canonically normalised ones. As we do not make assumptions about the scalar potential, let us expand the fields $\tau_i = \langle \tau_i \rangle + \delta \tau_i$ around the generic point ($\langle \tau_0\rangle$, $\langle \tau_1 \rangle$, $\langle \tau_2 \rangle$). The first entry fixes the value of the expansion parameter $\epsilon = 1/ \langle \tau_0 \rangle$, while the last entry is the classical value of the quintessence field. 
Eq. \eqref{eq:RotationGeneral} then gives the basis change matrix to be 
\begin{equation}
\mathcal{M} = \begin{pmatrix}
\frac{2}{\sqrt{3} \epsilon} && - \frac{2 \sqrt{6} \langle \tau_1 \rangle^{5/4}}{\gamma_1^{1/2}} \epsilon^{1/4} && - \frac{2 \sqrt{6} \langle \tau_2 \rangle^{5/4}}{\gamma_2^{1/2}} \epsilon^{1/4} \\
2 \sqrt{3} \langle \tau_1 \rangle && \frac{2 \sqrt{2} \langle \tau_1 \rangle^{1/4}}{\sqrt{3 \gamma_1} \epsilon^{3/4}} && \frac{2 \sqrt{6} \gamma_1 \gamma_2 \langle \tau_1 \rangle \langle \tau_2 \rangle^{5/4}}{\gamma_2 \langle \tau_1 \rangle^{1/2} - \gamma_1 \langle \tau_2 \rangle^{1/2}} \epsilon^{3/4} \\
2 \sqrt{3} \langle \tau_2 \rangle && \frac{2 \sqrt{6} \gamma_1 \gamma_2 \langle \tau_1 \rangle^{5/4} \langle \tau_2 \rangle}{\gamma_1 \langle \tau_2 \rangle^{1/2} - \gamma_2 \langle \tau_1 \rangle^{1/2}} \epsilon^{3/4} && \frac{2 \sqrt{2} \langle \tau_2 \rangle^{1/4}}{\sqrt{3 \gamma_2} \epsilon^{3/4}}
\end{pmatrix} \,,
\end{equation}
so that the moduli $\delta\tau_i$ can be written in terms of the canonically normalized fields $\varphi_0, \varphi_1, \varphi_2$ as
\begin{align}
\label{eq:BigfieldRotated}
\delta\tau_0 &= \mathcal{M}_{0i} \varphi_i = \frac{2}{\sqrt{3} \epsilon} \varphi_0 - \frac{2 \sqrt{6} \langle \tau_1 \rangle^{5/4}}{\gamma_1^{1/2}} \epsilon^{1/4} \varphi_1 - \frac{2 \sqrt{6} \langle \tau_2 \rangle^{5/4}}{\gamma_2^{1/2}} \epsilon^{1/4}  \varphi_2 \,, \\
\label{eq:SMfieldRotatedx}
\delta\tau_1 &= \mathcal{M}_{1i} \varphi_i = 2 \sqrt{3} \langle \tau_1 \rangle \varphi_0 + \frac{2 \sqrt{2} \langle \tau_1 \rangle^{1/4}}{\sqrt{3 \gamma_1} \epsilon^{3/4}} \varphi_1 + \frac{2 \sqrt{6} \gamma_1 \gamma_2 \langle \tau_1 \rangle \langle \tau_2 \rangle^{5/4}}{\gamma_2 \langle \tau_1 \rangle^{1/2} - \gamma_1 \langle \tau_2 \rangle^{1/2}} \epsilon^{3/4} \varphi_2 \,, \\
\label{eq:QfieldRotatedx}
\delta\tau_2 &= \mathcal{M}_{2i} \varphi_i = 2 \sqrt{3} \langle \tau_2 \rangle \varphi_0 + \frac{2 \sqrt{6} \gamma_1 \gamma_2 \langle \tau_1 \rangle^{5/4} \langle \tau_2 \rangle}{\gamma_1 \langle \tau_2 \rangle^{1/2} - \gamma_2 \langle \tau_1 \rangle^{1/2}} \epsilon^{3/4} \varphi_1 + \frac{2 \sqrt{2} \langle \tau_2 \rangle^{1/4}}{\sqrt{3 \gamma_2} \epsilon^{3/4}} \varphi_2 \,.
\end{align}
\noindent{{\underline{Couplings to Gauge Bosons}}}\\
For D7-branes wrapping $\tau_1$, the gauge coupling is determined by a holomorphic term
\begin{equation}
\label{gc}
\mathcal{L} \supset - \frac{\tau_1}{4 \pi} F^a_{\mu \nu} F^{a, \mu \nu} \,.
\end{equation}
Upon canonical normalisation of the fields\footnote{We use the  normalisation for the gauge fields in which their kinetic terms are given by $\mathcal{L} \supset - \frac{1}{4 e^2} F^a_{\mu \nu} F^{a, \mu \nu}$, as with this it is easier to compare with the bounds inferred in~\cite{Damour:2010rp}. }, the last term in Eq. \eqref{eq:SMfieldRotatedx} produces an effective dimension five coupling between the photon and the quintessence field 
\begin{equation}
\mathcal{L} \supset \frac{\sigma \varphi_2}{\langle\V\rangle^{1/2} } F^a_{\mu \nu} F^{a, \mu \nu} \,,
\end{equation}
where
\begin{equation}
\label{eq:Sigma}
\sigma = \frac{2 \sqrt{6} \gamma_1 \gamma_2 \langle \tau_1 \rangle \langle \tau_2 \rangle^{5/4}}{\gamma_2 \langle \tau_1 \rangle^{1/2} - \gamma_1 \langle \tau_2 \rangle^{1/2}} \,.
\end{equation}
On restoring units, the interaction is suppressed by a scale 
\be
\label{supr}
  \Lambda \sim M_{\rm pl} \V^{1/2}.
\ee
Note that the scale the suppression is weaker that $M_{\rm pl}$ by a factor of square root of the volume of the compactification.

\vspace{0.3  cm}

\noindent{{\underline{Couplings to Matter Fields}}}\\
Next, let us compute how the scalar $\tau_2$ couples to matter fields localised on a D7-brane wrapping the cycle $\tau_1$. For this, one requires a knowledge of the matter metrics in the visible sector. While these are not know in general, they can be determined in the limit of $\tau_0 \gg \tau_1$ ~\cite{Conlon:2006tj, Aparicio:2008wh, Aparicio:2015psl}. For matter arising from D7-branes wrapping the a blow-up cycle, the matter metric is:
\begin{equation}
K_{\alpha \beta} \sim \frac{\tau_1^{1/3}}{\V^{2/3}} \delta_{\alpha \beta} \simeq \frac{\tau_1^{1/3}}{\tau_0} \left(1 + \frac{2}{3} \frac{\tau_1^{3/2}}{\tau_0^{3/2}} + \frac{2}{3} \frac{\tau_2^{3/2}}{\tau_0^{3/2}}\right) \, \delta_{\alpha \beta} \,.
\end{equation}
Taking $\tau_i = \langle \tau_i \rangle + \delta \tau_i$, to leading order in the fluctuations
\begin{equation}
\label{matmet}
K_{\alpha \beta} = \frac{\langle \tau_1 \rangle^{1/3}}{\langle \V \rangle^{2/3}} \left(1 + \frac{1}{3} \frac{\delta \tau_1}{\langle \tau_1 \rangle} - \frac{\delta \tau_0}{\langle \tau_0 \rangle} + \frac{\gamma_1 \langle \tau_1 \rangle^{1/2}}{\langle \tau_0 \rangle^{3/2}} \delta \tau_1 + \frac{\gamma_2 \langle \tau_2 \rangle^{1/2}}{\langle \tau_0 \rangle^{3/2}} \delta \tau_2\right) \,,
\end{equation}
where $\langle \V \rangle = \alpha\left(\langle\tau_0\rangle^{3/2} - \gamma_1 \langle\tau_1\rangle^{3/2} - \gamma_2 \langle\tau_2\rangle^{3/2}\right)$.  The Lagrangian for matter fields, in particular the electron $e$ is given by
\begin{equation}
\mathcal{L} \supset K_{e \overline{e}} \overline{e} \gamma^\mu \partial_\mu e +  e^{K/2} \tilde{y} H \overline{e} e \,,
\end{equation}
where $\tilde{y}$ is the cubic superpotential Yukawa coupling. Note that
\begin{equation}
e^{K/2} \simeq  \V^{-1} \simeq \frac{1}{\langle \V \rangle} \left(1 - \frac{3}{2} \frac{\delta \tau_0}{\langle \tau_0 \rangle} + \frac{3}{2} \frac{\gamma_1 \langle \tau_1 \rangle^{1/2}}{\langle \tau_0 \rangle^{3/2}} \delta \tau_1 +  \frac{3}{2} \frac{ \gamma_2 \langle \tau_2 \rangle^{1/2}}{\langle \tau_0 \rangle^{3/2}} \delta \tau_2 + \mathcal{O}\left(\V^{-2}\right)\right) \,.
\end{equation}
Hence
\begin{align}
\mathcal{L} &\supset K_0 \overline{e} \gamma^\mu \partial_\mu e \left(1 + \frac{1}{3} \frac{\delta \tau_1}{\langle \tau_1 \rangle} - \frac{\delta \tau_0}{\langle \tau_0 \rangle} + \frac{\gamma_1 \langle \tau_1 \rangle^{1/2}}{\langle \tau_0 \rangle^{3/2}} \delta \tau_1 + \frac{\gamma_2 \langle \tau_2 \rangle^{1/2}}{\langle \tau_0 \rangle^{3/2}} \delta \tau_2\right) + \\
&+ \frac{\tilde{y} H}{\langle \V \rangle} \overline{e} e \left(1 - \frac{3}{2} \frac{\delta \tau_0}{\langle \tau_0 \rangle} + \frac{3}{2} \frac{\gamma_1 \langle \tau_1 \rangle^{1/2}}{\langle \tau_0 \rangle^{3/2}} \delta \tau_1 +  \frac{3}{2} \frac{\gamma_2 \langle \tau_2 \rangle^{1/2}}{\langle \tau_0 \rangle^{3/2}} \delta \tau_2\right) \,.
\end{align}
where $K_0 = \langle \tau_1 \rangle^{1/3}/\langle \V \rangle^{2/3}$. After canonical normalization of the electron field: $\hat{e} = \sqrt{K_0} e$ and electroweak symmetry breaking, the Lagrangian takes the form
\begin{align}
\mathcal{L} \supset \,\, & \overline{\hat{e}} \left(\gamma^\mu \partial_\mu + m_e\right) \hat{e} \left(1 +\frac{1}{3} \frac{\delta \tau_1}{\langle \tau_1 \rangle} - \frac{\delta \tau_0}{\langle \tau_0 \rangle} + \frac{\gamma_1 \langle \tau_1 \rangle^{1/2}}{\langle \tau_0 \rangle^{3/2}} \delta \tau_1 + \frac{\gamma_2 \langle \tau_2 \rangle^{1/2}}{\langle \tau_0 \rangle^{3/2}} \delta \tau_2\right) + \nonumber \\
\label{eq:couplings}
&- m_e \overline{\hat{e}} \hat{e} \left(\frac{1}{2} \frac{\delta \tau_0}{\langle \tau_0 \rangle} + \frac{1}{3} \frac{\delta \tau_1}{\langle \tau_1 \rangle} - \frac{1}{2} \frac{\gamma_1 \langle \tau_1 \rangle^{1/2}}{\langle \tau_0 \rangle^{3/2}} \delta \tau_1 - \frac{1}{2} \frac{\gamma_2 \langle \tau_2 \rangle^{1/2}}{\langle \tau_0 \rangle^{3/2}} \delta \tau_2\right) \,,
\end{align}
where $m_e$ is the electron mass. 
The contributions from the first line of Eq. \eqref{eq:couplings} vanish on-shell: the physical couplings are determined purely from the contributions in the second line. We can also neglect the term proportional to $\delta \tau_1/\langle \tau_0 \rangle^{3/2}$, since it is subleading with respect to the term proportional to $\delta\tau_1/\langle \tau_1\rangle$. Using the  expressions for the basis change in  (\ref{eq:BigfieldRotated}-\ref{eq:QfieldRotatedx}) we infer the couplings of the electron to the canonically normalised (quintessence) scalar $\varphi_2$:
\begin{equation}
\label{ecoup}
\mathcal{L}_{\rm int} \supset - \left(\frac{\sigma_0}{\langle \V \rangle^{5/6}} \frac{\varphi_2}{M_{\rm p}} + \frac{\sigma_1}{\langle \V \rangle^{1/2}} \frac{\varphi_2}{M_{\rm p}} + \frac{\sigma_2}{\langle \V \rangle^{1/2}} \frac{\varphi_2}{M_{\rm p}} \right) m_e \overline{\hat{e}} \hat{e} \,,
\end{equation}
where
\begin{equation}
\sigma_0 = \frac{\sqrt{6} \langle \tau_2 \rangle^{5/4}}{\gamma_2^{1/2}} \,, \qquad \sigma_1= \frac{2 \sqrt{2}}{\sqrt{3}}\frac{\gamma_1 \gamma_2 \langle \tau_2 \rangle^{5/4}}{\gamma_2 \langle \tau_1 \rangle^{1/2} - \gamma_1 \langle \tau_2 \rangle^{1/2}} \,, \qquad \sigma_2 = - \sqrt{\frac{2 \gamma_2}{3}} \langle \tau_2 \rangle^{3/4} \,.
\end{equation}
Notice that the electron coupling is suppressed by the same scale as in Eq. \pref{supr} for the gauge bosons, i.e. there is a factor of $\V^{1/2}$ in addition to the $M_{\rm pl}$ suppression of the dimension five operators\footnote{There are ambiguities in the form of the matter metric in Eq. \pref{matmet} at subleading order in the inverse volume expansion \cite{Blumenhagen:2009gk, Conlon:2011jq}. These can affect the last term in Eq. \pref{ecoup}, but the first and second term would not be affected.}. Couplings of the quarks can be computed in exactly the same manner, their interactions are also suppressed by the same scale.

\noindent{\underline{Bounds}}\\
Now, let us discuss implications of our result in the context of the bounds on fifth forces and time variation of fundamental constants. We have found $\varphi_2$ couples to both the gauge fields and Standard Model fermions with interactions which are suppressed by a scale $\Lambda \sim \V^{1/2} M_{\rm pl}$. As we want to infer order-of-magnitude contraints at the string scale, using a value of the gauge couplings compatible with unification $\alpha^{-1} \simeq 1/25$ (i.e. $4 e^2 \sim \mathcal{O}(1)$), the most stringent bound comes from
\begin{equation}
\label{eq:ExplicitConstraint}
d^c \simeq \frac{4 e^2 \sigma}{\V^{1/2}} < 10^{-6} \,,
\end{equation}
which translates into a lower bound on the volume:
 \begin{equation}
 \label{lower}
    \V > 10^{12},
    \end{equation}
which is a rather strong condition. In Eq. \eqref{eq:ExplicitConstraint} we have assumed that the blow-up dependent coefficient $\sigma$ in Eq. \eqref{eq:Sigma} is of $\mathcal{O}(1)$. Notice that this is a mild assumption and anyway a value of $\sigma$ which slightly deviates from $\mathcal{O}(1)$ does not affect the strong result in Eq. \eqref{lower}. The most stringent bounds on the time variation of fundamental constants are for the the fine structure constant \cite{Martins:2017yxk}, from the onset of domination of dark energy domination one requires
\begin{equation}
\label{alch}
{\delta \alpha \big{/} \alpha} < 10^{-6}.
\end{equation}
To compute the variation of $\alpha$ precisely one needs to track the  evolution of the quintessence field and carry of canonical normalisation along its trajectory\footnote{See \cite{Burgess:2010bz} for such a computation (in inverse volume expansion) in the inflationary context.}, but quick estimate can be made by assuming that the volume of the compactification remains approximately constant and the displacement of the blow-up mode is of the order of the string scale. This gives $\V  > 10^4$, which is much weaker than the condition obtained from the bounds on fifth forces in Eq. \pref{lower}.

\subsubsection{SM from D3-branes at a singularity}
\label{sec:SMSing}

Next, we analyse the case when the visible sector  degrees of freedom are located on D3-branes at a singularity; the geometrically separated light modulus will continue to be a blow-up. For D3-branes at a singularity, the gauge coupling is given by the dilaton; we shall be interested in the kinetic mixing of the the dilaton and the blow-up. Therefore, we begin by considering the \Kahler potential of the setup by including the universal $N=2$ supersymmetric  $\alpha'$ 
correction \cite{Becker:2002nn} as this induces the required mixing. The \Kahler potential is given by (see \cite{Conlon:2008wa}):
\begin{equation}
\label{singuni}
K = - 2 \log\left(\alpha \left(\tau_0^{3/2} - \gamma_2 \tau_2^{3/2}\right) + \frac{\xi s^{3/2}}{2}\right) - \log \left( S + \bar{S}\right) + \frac{\tau_1^2}{\V} \,,
\end{equation}
where $\tau_0$ is the universal \Kahler modulus, $S = s + i C_0$ the axio-dilaton, $\tau_1$ the modulus associated with the singularity and $\tau_2$ the geometric blow-up. At the singular locus, it is easy to see that $\tau_1$ does not kinetically mix with any other other moduli, the \Kahler metric in the $\{\tau_0, s, \tau_2\}$ subspace can be written as
$$
  K_{ij} =  K_{ij}^{(0)} +  K_{ij}^{(1)}  +  K_{ij}^{(2)}  + . . .
$$
with
\begin{equation}
K_{ij}^{(0)} = \begin{pmatrix}
A \epsilon^2 && 0 && 0 \\
0 && B  && 0 \\
0 && 0 && C \epsilon^{3/2}
\end{pmatrix} \,, \nonumber
\end{equation}
\begin{equation}
K_{ij}^{(1)} = \begin{pmatrix}
0 && D \epsilon^{5/2} && E \epsilon^{5/2} \\
D \epsilon^{5/2} && 0 && 0 \\
E \epsilon^{5/2} && 0 && 0 
\end{pmatrix} \,, \qquad K_{ij}^{(2)} = \begin{pmatrix}
0 && 0 && 0 \\
0 && 0 && F \epsilon^3 \\
0 && F \epsilon^3 && 0
\end{pmatrix} \,,
\end{equation}
where we have defined
\begin{equation}
A = \frac{3}{4} \,, \qquad B = \frac{1}{4 s^2} \,, \qquad C =  \frac{3 \gamma_2}{8 \sqrt{\tau_2}} \,, \nonumber
\end{equation}
\begin{equation}
D = \frac{9 \xi \sqrt{s}}{16 \alpha} \,, \qquad E = - \frac{9}{8} \gamma_2 \sqrt{\tau_2} \,, \qquad F = - \frac{9 \xi \gamma_2 \sqrt{s \tau_2}}{16 \alpha} \,.
\end{equation}
Using the same procedure as before, we get the basis change matrix to be:
\begin{equation}
\label{eq:RotationSMSing}
\mathcal{M} = \begin{pmatrix}
\frac{2}{\sqrt{3} \epsilon} && \frac{9 \xi}{2 \alpha} \langle s\rangle^{7/2} \epsilon^{5/2} && - \frac{2 \sqrt{6}}{\gamma_2^{1/2}} \langle\tau_2\rangle^{5/4} \epsilon^{1/4} \\
- \frac{3 \sqrt{3} \xi \langle s\rangle^{5/2}}{2 \alpha} \epsilon^{3/2} && 2 \langle s\rangle && \frac{3 \sqrt{3} \xi \gamma_2^{1/2}}{\sqrt{2} \alpha} \langle s\rangle^{5/2} \langle\tau_2\rangle^{3/4} \epsilon^{9/4} \\
2 \sqrt{3} \langle\tau_2\rangle && - \frac{9 \xi \gamma_2}{2 \alpha} \langle s\rangle^{7/2} \langle \tau_2\rangle^{1/2} \epsilon^{3} && \frac{2 \sqrt{2} \langle\tau_2\rangle^{1/4}}{\sqrt{3} \gamma_2^{1/2} \epsilon^{3/4}}
\end{pmatrix}\,.
\end{equation}

Note that the mixing of $s$ with $\tau_2$ (which determines the couplings of $\tau_2$ with gauge bosons) scales as $\epsilon^{9/4}$ which is much lower than the $\epsilon^{3/4}$ mixing that we had found for the corresponding quantity in the previous example (the mixing of $\tau_1$ with $\varphi_2$ in Eq. \pref{eq:SMfieldRotatedx}). Therefore, we turn to examining the case when the mixings arise from the inclusion of the leading loop corrections to the \Kahler potential in backgrounds with $N=1$ supersymmetry (instead of the universal $\alpha'$-correction considered above). The \Kahler potential takes the form:
\begin{equation}
\label{singloop}
K = - 2 \log\left(\alpha \left(\tau_0^{3/2} - \gamma_2 \tau_2^{3/2}\right)\right)   - \log \left( S + \bar{S}\right)+ \frac{\tau_1^2}{\V} + { {\hat{C}  \tau^{1/2}_{2}}\over {\left(S + \bar{S}\right) \tau_{0}^{3/2}} }.
\end{equation}
The last term in Eq. \pref{singloop} encodes the effect of loop corrections. Explicit computations in toroidal examples  and generic  effective field theory arguments \cite{Berg:2005ja, Conlon:2006gv, Berg:2007wt, Cicoli:2007xp, Cicoli:2008va} show that such a term is generated when a D7-branes wraps the cycle $\tau_2$. More generally, such a correction can be generated in $N=1$ backgrounds with NS-NS and R-R fluxes.\footnote{The loop we consider scales as $\alpha'^2$, it is an open question whether loop corrections at order $\alpha'$ exist or not. See \cite{Cicoli:2018kdo} for a detailed discussion of the present status of understanding of loop corrections in type IIB compactifications.} The kinetic matrix in the $\{\tau_0, s, \tau_2\}$ basis can be written as
 $$
     K_{ij} =  K_{ij}^{(0)} +  K_{ij}^{(1)}  +  K_{ij}^{(2)}  + . . .
 $$
with
\begin{equation}
K_{ij}^{(0)} = \begin{pmatrix}
A \epsilon^2  && 0 && 0 \\
 0 && B && 0  \\
0 && 0 && C \epsilon^{3/2}
\end{pmatrix} \,, \qquad
K_{ij}^{(1)} = \begin{pmatrix}
0 && 0 && 0 \\
 0 && 0 && F \epsilon^{3/2} \\
0 && F \epsilon^{3/2} && 0
\end{pmatrix} \,,
\end{equation}
\begin{equation}
K_{ij}^{(2)} = \begin{pmatrix}
0  && D \epsilon^{5/2} && E \epsilon^{5/2} \\
 D \epsilon^{5/2} && 0 && 0 \\
E \epsilon^{5/2} && 0 && 0
\end{pmatrix} \,,
\end{equation}
where we defined
\begin{equation}
A = \frac{3}{4} \,, \qquad B = \frac{1}{4 s^2} \,, \qquad C =  \frac{3 \gamma_2}{8 \sqrt{\tau_2}} \,, \nonumber
\end{equation}
\begin{equation}
D = {3 \hat{C} \sqrt{\tau_2} \over 16 s^2} \,, \qquad E = -{3 \hat{C} \over 32  s \sqrt{\tau_2}} - \frac{9 \gamma_2 \sqrt{\tau_2}}{8} \,, \qquad F = -{\hat{C} \over 16 s^2 \tau_2^{1/2}} \,.
\end{equation}
The basis change matrix takes the form
\begin{equation}
\label{eq:RotationLoopEffects}
\mathcal{M} = 
\begin{pmatrix}
\frac{2}{\sqrt{3} \epsilon} && \frac{3 \hat{C}}{2} \langle s \rangle \langle \tau_2 \rangle^{1/2} \epsilon^{5/2} && - \left(\frac{\hat{C}}{\sqrt{6} \gamma_2^{3/2}} \frac{\langle\tau_2\rangle^{1/4}}{\langle s \rangle} + \frac{2 \sqrt{6}}{\gamma_2^{1/2}} \langle\tau_2\rangle^{5/4}\right) \epsilon^{1/4} \\
-\frac{\sqrt{3} \hat{C}}{2} \langle \tau_2 \rangle^{1/2} \epsilon^{3/2} && 2 \langle s \rangle && \frac{\hat{C}}{\sqrt{6} \gamma_2^{1/2} \langle \tau_2 \rangle^{1/4}} \epsilon^{3/4}\\
2 \sqrt{3} \langle\tau_2\rangle + \frac{\hat{C}}{2 \sqrt{3} \gamma_2 \sqrt{\langle s\rangle}} && -\frac{\hat{C}}{2} \frac{\langle s\rangle}{\langle \tau_2 \rangle^{1/2}} \epsilon^{3/2} && \frac{2 \sqrt{2}}{\sqrt{3 \gamma_2}} \frac{\langle \tau_2\rangle^{1/4}}{\epsilon^{3/4}}
\end{pmatrix}\,.
\end{equation}

\noindent{\underline{{Couplings to Gauge Bosons}}}\\
In the case of SM at singularity, the gauge kinetic function is given by the axio-dilaton $S$, this translates to the coupling:
\begin{equation}
\label{eq:GaugeCouplingDil}
\mathcal{L} \supset  -\frac{s}{4} F^a_{\mu \nu} F^{a, \mu \nu} \,.
\end{equation}
In the case that the mixing is generated by the universal $\alpha'$ correction the basis change matrix in Eq. \eqref{eq:RotationSMSing} leads to a coupling between $\varphi_2$ and the gauge bosons 
\begin{equation}
\mathcal{L} \supset - \frac{\zeta \langle s\rangle^{5/2}}{\langle\V\rangle^{3/2}}\, \varphi_2 F^a_{\mu \nu} F^{a,\,\mu \nu} \,,
\end{equation}
where
\begin{equation}
\zeta = \frac{3 \sqrt{3} \xi \gamma_2^{1/2}}{4 \sqrt{2} \alpha} \langle\tau_2\rangle^{3/4} \,,
\end{equation}
On restoring units, this corresponds on a suppression scale 
%
\be
\label{supr}
  \Lambda \sim M_{\rm pl} \frac{\langle\V\rangle^{3/2}}{\langle s \rangle^{5/2}}.
\ee
On the other hand, if the loops effects are responsible for the mixing, the basis change matrix in Eq. \eqref{eq:RotationLoopEffects} gives the coupling between $\varphi_2$ and the gauge fields to be
\begin{equation}
\mathcal{L} \supset \frac{\zeta}{\langle\V\rangle^{1/2}} \, \varphi_2 F^a_{\mu \nu} F^{a,\,\mu \nu} \,,
\end{equation}
where
\begin{equation}
\zeta = \frac{\hat{C}}{\sqrt{6} \gamma_2^{1/2} \langle \tau_2 \rangle^{1/4}} \,,
\end{equation}
and $\varphi_2$ is the canonically normalized quintessence field. On restoring units, this corresponds to a suppression scale
\be
\label{supr}
  \Lambda \sim M_{\rm pl} \langle\V\rangle^{1/2}.
\ee
\noindent{{\underline{Coupling to Matter Fields}}}\\
For matter fields arising from D3-branes at singularities, in the limit $\tau_0 \gg 1$, the \Kahler matter metric is given by $K_{\alpha \beta} = \V^{-2/3}$ \cite{Conlon:2006tj, Aparicio:2008wh, Aparicio:2015psl}. On expanding the fields as $\tau_i = \langle \tau_i \rangle + \delta \tau_i$ one finds:
\begin{equation}
K_{\alpha \beta} = \frac{\delta_{\alpha \beta}}{\langle \V \rangle^{2/3}} \left(1  - \frac{\delta \tau_0}{\langle \tau_0 \rangle} + \frac{\gamma_2 \langle \tau_2 \rangle^{1/2}}{\langle \tau_0 \rangle^{3/2}} \delta \tau_2\right) \,,
\end{equation}
where $\langle \V \rangle = \alpha \left(\langle \tau_0 \rangle^{3/2} - \gamma_2 \langle \tau_2 \rangle^{3/2} \right)$.  This give the interaction term involving the electrons can be computed as
before, it turns out to be:
\begin{align}
\label{eq:CouplingElectronSing}
\mathcal{L} \supset - m_e \overline{\hat{e}} \hat{e} \left(\frac{1}{2} \frac{\delta \tau_0}{\langle \tau_0 \rangle} - \frac{1}{2} \frac{\gamma_2 \langle \tau_2 \rangle^{1/2}}{\langle \tau_0 \rangle^{3/2}} \delta \tau_2\right) \,,
\end{align}
where $m_e$ the electron mass. In the case that the mixing arises from the universal $\alpha'$ correction (the corresponding basis change is given by Eq. \eqref{eq:RotationSMSing}) then the coupling between electrons and $\varphi_2$ is

\begin{equation}
\mathcal{L}_{\rm int} \supset \left(\frac{\sigma_0}{\langle \V \rangle^{5/6}} \frac{\varphi_2}{M_{\rm p}} + \frac{\sigma_2}{\langle \V \rangle^{1/2}} \frac{\varphi_2}{M_{\rm p}} \right) m_e \overline{\hat{e}} \hat{e} \,,
\end{equation}
where
\begin{equation}
\sigma_0 = -\frac{\sqrt{6}}{\gamma_2^{1/2}}  \langle \tau_2 \rangle^{5/4} \,, \qquad \sigma_2 = \sqrt{\frac{2}{3}} \gamma_2^{1/2} \langle \tau_2 \rangle^{3/4} \,.
\end{equation}
On the other hand, if the loop effects are relevant (and the corresponding basis change matrix is in Eq. \eqref{eq:RotationLoopEffects}) then the coupling between electrons
and $\varphi_2$ is
\begin{equation}
\mathcal{L}_{\rm int} \supset - \left(\frac{\sigma_0}{\langle \V \rangle^{5/6}} \frac{\varphi_2}{M_{\rm p}} - \frac{\sigma_2}{\langle \V \rangle^{1/2}} \frac{\varphi_2}{M_{\rm p}} \right) m_e \overline{\hat{e}} \hat{e} \,,
\end{equation}
where
\begin{equation}
\sigma_0 = \left(\frac{\hat{C}}{2 \sqrt{6} \gamma_2^{3/2}} \frac{\langle\tau_2\rangle^{1/4}}{\langle s \rangle} + \frac{\sqrt{6}}{\gamma_2^{1/2}} \langle\tau_2\rangle^{5/4}\right) \,, \qquad \sigma_2 = \sqrt{\frac{2}{3}} \gamma_2^{1/2} \langle \tau_2 \rangle^{3/4} \,.
\end{equation}

Before discussing the lower bounds on the volume that the above interactions imply, we would like to note that in the case that the mixings are induced by the universal $\alpha'$-correction, the field $\varphi_2$ couples to the gauge bosons and matter fermions with different strengths. The reason for this is simple: its coupling to the gauge bosons is determined by the mixing of $\tau_2$ to the dilaton, while its couplings to the matter fields are given by its mixings with the volume modulus. As discussed in Section \ref{sec:fifth}, the detailed phenomenology of the implications of such non-universal couplings for violations of the equivalence principle is yet to be developed. Our results give motivation for such an analysis.

\vspace{0.2 cm}

\noindent \underline{Bounds}\\
As mentioned above, in the case that the mixings are generated by the $\alpha'$ corrections the couplings are non-universal. It is natural to expect that the coupling to the gluons are the most relevant for the bounds \cite{Adelberger:2003zx} (they are also weaker than the couplings of matter fields in the case at hand).  Requiring that the gluon couplings satisfy the condition in Section \ref{sec:fifth}, i.e. $(d^c_{g})^2  < 10^{-12}$, one finds
\begin{equation}
  \V \gtrsim 10^4/g_s^{5/3} \simeq 5 \times 10^5\,, \quad \text{in the case of mixing from the universal $\alpha'$-corrections} \,,
\end{equation}
where in the last step we have taken $g_s \simeq 0.1$. On the other hand, in the case that the mixing arises from the leading loop corrections we obtain the following constraint
\begin{equation}
  \V \gtrsim 10^{12}.  \quad  
\end{equation}
Both bounds are strong, but our results illustrate the importance of having detailed knowledge of the structure of quantum corrections in any model for addressing fifth force bounds
 
\subsubsection{M-theory with single blow-up}
Next, we consider an example in M-theory compactified on a manifold of $G_2$ holonomy. The M-theory \Kahler potential is proportional to the $\log$ of the volume of the compactification expressed in terms of the three cycles volumes. For our purposes it will be sufficient to consider a toy example with two moduli - the volume $\phi_0$ and one blow-up $\phi_1$, with $\langle \phi_0 \rangle \gg \langle \phi_1 \rangle$. The modulus $\phi_1$ plays the role of the quintessence field, while $\phi_0$ sets the gauge coupling $\alpha_{\rm GUT}$. The \Kahler potential takes the form
$$
  K =  -3 \log \left(  \left( {   s_0 + \bar{s}_0 \over 2 } \right)^{7/3} -    \left(  {s_1 + \bar{s}_1  \over 2 } \right)^{7/3}  \right)  \,,
$$
where $\phi_0 = \rm{Re}(s_0)$ and $\phi_1 = \rm{Re}(s_1)$. The kinetic matrix in the limit $\phi_0 \gg \phi_1$ is
\begin{equation}
K_{ij} = \begin{pmatrix}
{ 7  \over 4 \phi_0^{2} }&& -{ 49 \phi_1^{4/3} \over 12 \phi_0^{10/3}}&&  \\
  -{ 49 \phi_1^{4/3} \over 12 \phi_0^{10/3}} &&{7 \phi_1^{1/3}\over 3 \phi_0^{7/3}}
\end{pmatrix} \,,
\end{equation}%
the basis change matrix is easily obtained:
\begin{equation}
\mathcal{M} = \begin{pmatrix}
\frac{2 \langle \phi_0 \rangle}{\sqrt{7}} & \sqrt{\frac{7}{3}} \frac{\langle \phi_1 \rangle^{7/6}}{\langle \phi_0 \rangle^{1/6}} \\
- \frac{2 \sqrt{7} \langle \phi_1 \rangle^{4/3}}{3 \langle \phi_0 \rangle^{1/3}} & \sqrt{\frac{3}{7}} \frac{\langle \phi_0 \rangle^{7/6}}{\langle \phi_1 \rangle^{1/6}} \,.
\end{pmatrix}
\end{equation} 

\noindent{{\underline{Couplings to Gauge Bosons}}}\\
The gauge coupling is determined by a holomorphic term
\begin{equation}
\label{gc}
\mathcal{L} \supset - \frac{\phi_0}{4} F^a_{\mu \nu} F^{a, \mu \nu} \,.
\end{equation}
Upon canonical normalisation of the fields, the coupling between the photon and $\varphi_2$ is 
\begin{equation}
\label{eq:MTheoryGaugeCoupling}
\mathcal{L} \supset \frac{\sigma \varphi_1}{\langle\V\rangle^{1/14} } F^a_{\mu \nu} F^{a, \mu \nu} \,,
\end{equation}
where
\begin{equation}
\label{eq:Sigma}
\sigma = - \frac{1}{4} \sqrt{\frac{7}{3}} \langle \phi_1 \rangle^{7/6}  \,.
\end{equation}
On restoring units, the interaction is suppressed by a scale 
\be
\label{mthsupr}
  \Lambda \sim M_{\rm pl} \V^{1/14}.
\ee
Given the low power of $\V$ in Eq. \pref{mthsupr}, $\varphi_2$ has essentially Planck suppressed couplings with gluons for realistic value of the volume; thus cannot be very light.

%

\subsection{Fibre Models}

A large class of Calabi-Yaus are fibrations. Fibre moduli can have weaker-than-Planck suppressed interactions with open string degrees of freedom that are localised in the compactification, hence are interesting candidates for the being the quintessence scalar\footnote{The model in \cite{Cicoli:2012tz} uses a fibre modulus as the quintessence field. The model relies on the supersymmetric large extra-dimensions (SLED) proposal \cite{Aghababaie:2003wz}. See  \cite{Cicoli:2011yy} for a discussion of embedding of SLED in string theory and the associated challenges.}. 
 
\subsubsection{Standard Model at a geometric blow-up}
 
We begin by analysing the case where the light field is a fibre modulus and the visible sector is realised by a blow-up mode in the geometric regime. We will consider the simplest constructions where the \Kahler potential takes the form\footnote{For realisations in explicit Calabi-Yaus see \cite{Cicoli:2011it}.}:
\begin{equation}
\label{fib}
  K = -2 \log \V = - 2 \log \left(\hat{\V}  -  \gamma \tau_1^{3/2}\right) \,  \phantom{abc} { \text{with}}  \phantom{abc} \hat{\V}  = \alpha \sqrt{\tau_2} \left( \tau_0 - \beta \tau_2 \right) \,,
\end{equation}
where $\tau_0$ is the volume of the base, $\tau_1$ volume of the fibre, $\tau_2$ is the volume of the blow-up mode, and $\alpha, \beta$ and $\gamma$ are constants. For simplicity, in the following we will consider the case $\beta = 0$ (see \cite{Burgess:2016owb} for a recent discussion on such models in the context of fibre inflation). In the large volume limit $\tau_{0} \gg \tau_1, \tau_2$ the volume of the internal manifold is approximately: 
\begin{equation}
\V \simeq \alpha \sqrt{\tau_2} \tau_0 \ . 
\end{equation}
Using the basis $\{\tau_2, \V, \tau_1\}$ the kinetic matrix can be written as
\begin{equation}
K_{ij} = K^{(0)}_{ij} + K^{(1)}_{ij} \,,
\end{equation}
 where $\delta = \V^{-1}$ and
\begin{equation}
K_{ij}^{(0)} = \begin{pmatrix}
A & 0 & 0 \\
0 & B \delta^2 & 0 \\
0 & 0 & C \delta
\end{pmatrix} \,, \qquad
K_{ij}^{(1)} = \begin{pmatrix}
0 & D \delta & E \delta \\
D \delta & 0 & 0 \\
E \delta & 0 & 0
\end{pmatrix} \,,
\end{equation}
and we have defined
\begin{equation}
A = \frac{3}{8 \tau_2^2} \,, \qquad B = \frac{1}{2} \,, \qquad C = \frac{3 \alpha \gamma}{8 \sqrt{\tau_1}} \,, \qquad D = - \frac{1}{4 \tau_2} \,, \qquad E = - \frac{3 \alpha \gamma \sqrt{\tau_1}}{8 \tau_2} \,.
\end{equation}
The basis change matrix is given by
\begin{equation}
\label{eq:RotationFibre}
\mathcal{M} = \begin{pmatrix}
2 \sqrt{\frac{2}{3}} \langle\tau_2\rangle & \frac{2 \sqrt{2}}{3} \langle \tau_2 \rangle & 2 \sqrt{\frac{2 \alpha \gamma}{3}} \langle \tau_1 \rangle^{3/4} \langle \tau_2 \rangle \delta^{1/2} \\
- \frac{4}{3} \sqrt{\frac{2}{3}} \langle \tau_2 \rangle^2 \delta & \frac{\sqrt{2}}{\delta} & - \frac{4}{3} \sqrt{\frac{2}{3 \alpha \gamma}} \langle \tau_1 \rangle^{5/4} \delta^{1/2}  \\
- 2 \alpha \gamma \sqrt{\frac{2}{3}} \langle \tau_1 \rangle^{1/2} \langle \tau_2 \rangle^2 \delta & \frac{2 \sqrt{2}}{3} \langle \tau_1 \rangle & \frac{2 \sqrt{2} \langle\tau_1\rangle^{1/4}}{\sqrt{3 \alpha \gamma}} \frac{1}{\delta^{1/2}}
\end{pmatrix} \,.
\end{equation}

\noindent{{\underline{Couplings to Gauge Bosons}}}\\ From the basis change matrix in Eq. \eqref{eq:RotationFibre} and using Eq. \eqref{gc} it is easy to see that the coupling between the gauge bosons and the quintessence field is given by
\begin{equation}
\mathcal{L} \supset \frac{\zeta \varphi_2}{\langle\V\rangle} F^a_{\mu \nu} F^{a,\,\mu \nu} \,,
\end{equation}
where
\begin{equation}
\zeta = - 2 \alpha \gamma \sqrt{\frac{2}{3}} \langle \tau_1 \rangle^{1/2} \langle \tau_2 \rangle^2 \,.
\end{equation}
On restoring units, the interaction is suppressed by a scale 
\be
\label{supr}
  \Lambda \sim M_{\rm pl} \langle\V\rangle.
\ee
\noindent{\underline{Couplings to Matter Fields}}\\ 
The effective coupling of electrons changes slightly with respect to the previous Section due to the different expression for the volume
\begin{equation}
\mathcal{L}_{\rm int} \supset m_e \overline{\hat{e}} \hat{e} \left(-\frac{1}{3} \frac{\delta \V}{\langle \V \rangle} - \frac{1}{3} \frac{\delta \tau_1}{\langle \tau_1 \rangle}\right) \,,
\end{equation}
which, after the basis change leads to 
\begin{equation}
\mathcal{L}_{\rm int} \supset \left(\frac{\sigma_1}{\langle \V \rangle^2} \frac{\varphi_2}{M_{\rm p}} + \frac{\sigma_2}{\langle \V \rangle} \frac{\varphi_2}{M_{\rm p}} \right) m_e \overline{\hat{e}} \hat{e} \,,
\end{equation}
where
\begin{equation}
\sigma_1 = \frac{4}{9} \sqrt{\frac{2}{3}} \langle \tau_2 \rangle^2 \,, \qquad \sigma_2 = \frac{2 \alpha \gamma}{3} \sqrt{\frac{2}{3}} \frac{\langle \tau_2 \rangle^2}{\langle \tau_1 \rangle^{1/2}} \,.
\end{equation}
\noindent{\underline{Bounds}}\\ 
The bounds inferred by requiring that $\left(d_g^c\right)^2 < 10^{-12}$ is
\begin{equation}
\V \gtrsim 10^6 \,.
\end{equation}

\subsubsection{Fibre Models with SM at Singularity}

Next, let us consider the case in which the light field continues to be a fibre modulus, but the visible sector fields are realised by branes at singularities. Incorporating the effects of the universal $\alpha'$-correction, the \Kahler potential takes the form
\begin{equation}
\label{fib}
  K = - 2 \log \left( {\V} + \frac{\xi s^{3/2}}{2}\right) - \log\left(2s\right)  + \frac{\tau_1^2}{\V} \,,
\end{equation}
with the volume as in Eq. \eqref{fib}. In the basis $\{\tau_2, \V, s\}$, the kinetic matrix can be written as
$$
  K_{ij} = K_{ij}^{(0)} + K_{ij}^{(1)} + K_{ij}^{(2)} + ...
$$
with
\begin{equation}
K_{ij}^{(0)} = 
\begin{pmatrix}
A & 0 & 0 \\
0 & B \delta^2 & 0 \\
0 & 0 & C
\end{pmatrix} \,, \qquad
K_{ij}^{(1)} = \begin{pmatrix}
0 & D \delta & 0 \\
D \delta & 0 & 0 \\
0 & 0 & 0
\end{pmatrix} \,, \qquad
K_{ij}^{(2)} = 
\begin{pmatrix}
0 &  0 & 0 \\
0 & 0 & F \delta^2 \\
0 & F \delta^2 & 0
\end{pmatrix} \,,
\end{equation}
where
\begin{equation}
A = \frac{3}{8 \tau_2^2} \,, \qquad B = \frac{1}{2} \,, \qquad C = \frac{1}{4 s^2} \,, \qquad D = - \frac{1}{4 \tau_2} \,, \qquad F = \frac{3 \xi \sqrt{s}}{8} \,.
\end{equation}
The basis change matrix is given by
\begin{equation}
\label{eq:RotationFibreSing}
\mathcal{M} = \begin{pmatrix}
2 \sqrt{\frac{2}{3}} \langle \tau_2 \rangle && \frac{2 \sqrt{2}}{3} \langle \tau_2 \rangle && \frac{6 \xi \langle s \rangle^{11/2} \langle \tau_2 \rangle}{3 \langle s \rangle^2 - 2 \langle \tau_2 \rangle^2} \delta^3 \\
-\frac{4}{3} \sqrt{\frac{2}{3}} \langle \tau_2 \rangle^2 \delta && \frac{\sqrt{2}}{\delta} && 3 \xi \langle s \rangle^{7/2} \delta^2 \\
\frac{4 \sqrt{2}}{\sqrt{3}} \frac{\xi \langle s \rangle^{5/2} \langle \tau_2 \rangle^4}{2 \langle \tau_2 \rangle^2 - 3 \langle s \rangle^2} \delta^3 && - \frac{3}{\sqrt{2}} \xi \langle s \rangle^{5/2} \delta && 2 \langle s \rangle
\end{pmatrix}\,.
\end{equation}
Note that the mixings induced between the dilaton and the fibre is rather small (it scales as $\delta^{3}$), this it is important to consider loop effects. Following \cite{Berg:2005ja, Conlon:2006gv, Berg:2007wt, Cicoli:2007xp, Cicoli:2008va}, we take the \Kahler potential to be\footnote{As in Section \label{sec:SMSing} we take the loop correction to be scaling as $\alpha'^{2}$. A correction scaling as $\alpha'$ if present will lead to stronger mixings, thus our results can be be considered as lower bounds.} 
\begin{equation}
K = - 2 \log \hat{\V} - \log\left(2s\right) - \frac{\lambda_2}{s \tau_2} \,.
\end{equation}
 Using the basis $\{\tau_2, \V, s\}$, in the regime $\tau_0 \gg \tau_f \gg 1$, the kinetic matrix can be written as
\begin{equation}
K_{ij} = K^{(0)}_{ij} + \epsilon_1 K^{(1)}_{ij} + \epsilon_2 K^{(2)}_{ij} \,,
\end{equation}
where $\epsilon_1 = \frac{1}{4 \tau_2^2 s^2}$, $\epsilon_2 = \V^{-1}$ and $\epsilon_2 \ll \epsilon_1$. With
\begin{equation}
K^{(0)}_{ij} = \begin{pmatrix}
A & 0 & 0 \\
0 & B \epsilon_2^2 & 0  \\
0 & 0 & C
\end{pmatrix} \,, \quad
K^{(1)}_{ij} = \begin{pmatrix}
0 & 0 & E \epsilon_1 \\
0 & 0 & 0  \\
E \epsilon_1 & 0 & 0
\end{pmatrix} \,, \quad
K^{(2)}_{ij} = \begin{pmatrix}
0 & D \epsilon_2 & 0 \\
D \epsilon_2 & 0 & 0 \\
0 & 0 & 0
\end{pmatrix} \,,
\end{equation}
where we have defined
\begin{equation}
A = \frac{3}{8 \tau_2^2} \,, \quad B = \frac{1}{2} \,, \quad C = \frac{1}{4 s^2} \,, \quad D = -\frac{1}{4 \tau_2} \quad E = -\lambda_2 \,.
\end{equation}
The basis change matrix is given by
\begin{equation}
\label{eq:RotationFibreLoop}
{\cal M} =
\begin{pmatrix}
{2 \sqrt{2}  \tau_2 \over \sqrt{3} } &&  { 2 \sqrt{2} \tau_2 \over 3 } && {16 \lambda_2 \tau_2^2 s^3 \epsilon_1 \over {3s^2 - 2 \tau_2^2}}\\
-{ 4 \sqrt{2} \tau_2^2 \epsilon_2 \over 3 \sqrt{3} } && { \sqrt{2} \over \epsilon_2}  && - {16 \lambda_2 \tau_2 s^5 \epsilon_1 \epsilon_2 \over {3s^2 - 2 \tau_2^2}}\\
{16 \sqrt{2} \lambda_2 \tau_2^3 s^{2} \epsilon_1 \over {\sqrt{3} (2 \tau_2^2  - 3s^2)}}  && {2 \sqrt{2} \lambda_2 \over 3 \tau_2} && { 2s} \\
\end{pmatrix}
\end{equation}
Next, let us compute the couplings of $\varphi_2$ to gauge bosons and matter fields.

\noindent{{\underline{Couplings to Gauge Bosons}}}\\  In the case that the mixings are generated by the universal $\alpha'$ corrections, it is easy to see from the basis change matrix in Eq. \eqref{eq:RotationFibreSing} and Eq. \eqref{eq:GaugeCouplingDil} that the coupling between the gauge bosons and $\varphi_2$ is given by
\begin{equation}
\mathcal{L} \supset \frac{\zeta \varphi_2}{\langle\V\rangle^3} F^a_{\mu \nu} F^{a,\,\mu \nu} \,,
\end{equation}
where
\begin{equation}
\zeta = - \sqrt{\frac{2}{3}} \frac{\xi \langle s \rangle^{1/2} \langle \tau_2 \rangle^4}{3 - 2 \left(\frac{\langle \tau_2 \rangle}{\langle s \rangle}\right)^2} \,.
\end{equation}
Restoring units, the interaction is suppressed by the scale 
\be
\label{supr}
  \Lambda \sim M_{\rm pl} \langle\V\rangle^3.
\ee
On the other hand, if the mixing is due to loop effects, as given in Eq. \eqref{eq:RotationFibreLoop}, then the coupling takes the form
\begin{equation}
\label{eq:CouplingsLoopFibre}
\mathcal{L} \supset \zeta \varphi_2 F^a_{\mu \nu} F^{a,\,\mu \nu} \,,
\end{equation}
where
\begin{equation}
\label{lmixdanger}
\zeta = - {4 \sqrt{2} \lambda_2 \tau_2^3 s^{2} \epsilon_1 \over {\sqrt{3} (2 \tau_2^2  - 3s^2)}} \ .
\end{equation}
\noindent{\underline{Couplings to Matter Fields}}\\ 
The couplings to electrons and other matter fermions can computed as in Section \ref{sec:SMSing}. One finds
  \begin{equation}
\mathcal{L}_{\rm int} \supset m_e \overline{\hat{e}} \hat{e} \left(-\frac{1}{3} \frac{\delta \V}{\langle \V \rangle}\right) \,.
\end{equation}
After making use of the  basis change matrices in Eq. \eqref{eq:RotationFibreSing} and \label{eq:RotationFibreLoop} this leads couplings
\begin{equation}
\label{eq:MatterCouplingFibreSing}
\mathcal{L}_{\rm int} \supset \frac{\sigma_1}{\langle \V \rangle^2} \frac{\varphi_2}{M_{\rm p}} \, m_e \overline{\hat{e}} \hat{e} \,,
\end{equation}
where
\begin{equation}
\label{fsuni}
\sigma_1 = \frac{4}{9} \sqrt{\frac{2}{3}} \langle \tau_2 \rangle^2 \,.
\end{equation}
for both the cases.

\noindent{\underline{Bounds}}\\ 
As in Sec. \ref{sec:SMSing}, the couplings to matter and to gauge bosons have different strengths. As discussed earlier, a detailed
analysis of the bounds for such cases is yet to be done. In the case that the mixings are generated by the universal $\alpha'$ correction; imposing that the condition $\left(d_g^c\right)^2 < 10^{-12}$ (as one expects the gluon couplings to be most relevant for the bounds\footnote{The gluon couplings are also weaker than the matter couplings for the case at hand, this the condition used is conservative.}), one finds $\V \gtrsim 10^2 $. The condition is not very strong. Thus it is important to check the effect that loops  have. From Eq. \pref{lmixdanger}, one sees that in at generic points in the moduli space the loop effects general a coupling which has no volume suppression, preventing $\varphi_2$ from light. For large $\tau_2$ the coupling scales as $\tau_2^{-1}$, even in the case $\tau_2 \sim \V^{2/3} \gg 1$, one would obtain strong bounds.

\vspace{0.5cm}

Before closing this Section we would like to note that one can compare the strength of mixings obtained by us to those obtained by diagonalising both the kinetic and mass matrices \cite{Conlon:2007gk, Cicoli:2010ha} (which were done in the presence of a specific potential). In all cases, the mixings obtained by us are of lower or equal strength. This is in keeping with the expectation that our results should be considered as lower bounds on the strength of the interactions.

\section{Kinetic Mixing of $U(1)$ fields:}
\label{sec:uone}

In this Section, we will consider the situation where the quintessence scalar does not couple directly to the Standard Model photon, but
has direct couplings to a hidden sector photon and analyse the implications that kinetic mixing between the $U(1)s$ has for quintessence. We take the tree level Lagrangian to be of the form:

$$
  \mathcal{L}  \supset  - {1 \over 4 e^2} F^{1}_{\mu \nu} F^{1 \,\mu \nu} -  {1 \over 4 } h \left({ \phi / M_{\rm pl}}\right)F^2_{\mu \nu} F^{2 \, \mu \nu}
$$
where $F_{\mu \nu}^{1}$ is the electromagnetic field strength and $F_{\mu \nu}^{2}$ is the field strength of associated with a gauge field which is in the sector of the quintessence field $(\phi)$. We will confine our analysis to the case where there is a single hidden photon, the arguments easily generalise to cases with multiple hidden photons. While geometric separation naturally leads to such a structure in the tree level kinetic terms, but this is not preserved once quantum effects are incorporated. Integrating out heavy bi-fundamental string states leads to kinetic mixing between gauge fields \cite{Holdom:1985ag, Dienes:1996zr, Abel:2008ai}. In general, if the gauge couplings of the 
 two sectors are $g_{1}$ and $g_2$ generic estimates of the loop factors give the strength of the kinetic mixing to be
\bel{strmix}
    \lambda \approx {1 \over 12 \pi^2 }g_1 g_2.
\ee
Incorporating the couplings to matter, the Lagrangian takes the form
\bel{matcoup}
    {\cal L} \supset - \frac14 \, Z_{ab} F^a_{\mu\nu} F^{b\,\mu\nu}
    - {1 \over 2} M^2_{ab} A^a_{\mu} A^{b\,\mu} - j^\mu_a A^a_\mu \,,
\ee
where $F^a_{\mu\nu} = \partial_\mu A^a_\nu - \partial_\nu A^a_\mu$
and $j^\mu_a$  are the currents to which each gauge boson
couples. We take the kinetic and mass matrices to be of the form
\bel{matri}
    Z = \left(%
    \begin{array}{ccc}
    1 && \lambda \\
    \lambda && 1 \\
    \end{array}%
    \right) \qquad \hbox{and}\qquad
    M^2 = \left(%
    \begin{array}{cc}
    \ma^2 & \mu^2 \\
    \mu^2 & \mb^2 \\
    \end{array}%
    \right)
\ee
Diagonalising the kinetic and mass matrices, the Lagrangian becomes:
\bel{lfinal}
    {\cal L} = -\frac14 \, {\cal F}^a_{\mu\nu} {\cal F}^{\mu\nu}_a
    - \frac12 \, \left( M_+^2 {\cal A}^+_\mu {\cal A}^\mu_+
    + M_-^2 {\cal A}^-_\mu {\cal A}_-^\mu \right)
    + {N^a}_b j^\mu_a {\cal A}^b_\mu \,,
\ee
where 
\bel{nmatrix}
    N = \frac{1}{\sqrt{1-\lambda^2}} \left(%
    \begin{array}{ccc}
    \cos(\theta+\halpha) && -\sin(\theta + \halpha) \\
    \sin(\theta-\halpha) && \cos(\theta - \halpha) \\
    \end{array}%
    \right) \,.
\ee
where the angles $\theta$ and $\halpha$ are 
\be
    \sin 2\halpha = \lambda \qquad \hbox{and} \qquad
    \tan 2\theta = \frac{2\mu^2 - (\ma^2 + \mb^2)
    \lambda}{(\ma^2 - \mb^2) \sqrt{1-\lambda^2}} \,.
\ee
The mass eigenvalues $M_{\pm}$ are
\be
    M^2_{\pm} = \frac{\ma^2 + \mb^2 - 2 \lambda \mu^2
    \pm \Delta}{2(1 - \lambda^2)} \,,
\ee
with
\be
    \Delta^2 = (\ma^2 - \mb^2)^2 + 4\mu^4 - 4\lambda \mu^2 (\ma^2
    + \mb^2) + 4\lambda^2 \ma^2 \mb^2 \,.
\ee

Now, let us analyse the implications this has for time dependence of couplings in the Standard Model sector.

\begin{itemize}

\item If $\ma^2 = \mu^2 = 0$, then one of the gauge fields  (corresponding to the photon) is
massless. The other gauge field has mass $M_+^2 =
\mb^2/(1-\lambda^2)$ . Also $\sin 2\theta = \sin 2\halpha = \lambda$, as a result of this only the massive (hidden) vector acquires a coupling to both currents
\be \label{eq:masslessmassivemixing}
    {\cal L}_{\rm int} = j_1^\mu {\cal
    A}^-_\mu + \frac{1}{\sqrt{1-\lambda^2}}
    \left( j_2^\mu - \lambda
    j_1^\mu \right) {\cal A}^+_\mu \,.
\ee
The couplings of the massless photon are unaffected by the mixing. Hence, a time variation of the hidden sector gauge coupling due to a rolling quintessence scalar does not affect the fine structure constant.

\item Next let us  $M^2_{ab} = 0$. In this case,  one can take $\theta = 0$,  $N = Z^{-1/2}$;
both gauge bosons in general couple to both currents
\be\label{eq:masslessBoson}
    {\cal L}_{\rm int} = \frac{\cos\halpha}{\sqrt{1-\lambda^2}}
    \left( j_1^\mu {\cal A}^-_\mu
    + j_2^\mu {\cal A}^+_\mu \right)
    - \frac{\sin\halpha}{\sqrt{1-\lambda^2}}
    \left( j_1^\mu {\cal A}^+_\mu
    + j_2^\mu {\cal A}^-_\mu \right) \,,
\ee
with $\sin^2\halpha = \frac12 \left[ 1 - \sqrt{1-\lambda^2}
\right]$ and $\cos^2\halpha = \frac12 \left[ 1 + \sqrt{1 -
\lambda^2} \right]$ .
Note that, in this case the coupling of the visible sector photon ${\cal{A}}_{\mu}^{1}$  to $j^{\mu}_1$ {\it depends} on the 
mixing parameter. Thus, a change in the mixing parameter caused due to a rolling quintessence scalar leads to time variation in the fine structure constant. For small $\lambda$, $\cos \halpha \sim 1 - {\lambda^2 \over 8}$. Thus we have $ { \delta \alpha / \alpha} =  {\delta(\lambda^2) \over 4}$. Making use of the expression for for strength of the mixing parameter in Eq. \pref{strmix} 
$$
  { \delta \alpha \over \alpha} \simeq  {\alpha \delta( g^2_{\rm hidden} ) \over  144 \pi^3}
$$
Note that even for an ${\mathcal{O}}(1)$ differential variation of the hidden sector coupling, the differential variation in the fine structure constant is within the bounds ${\delta \alpha / \alpha} <  10^{-6} $.
\end{itemize}

In summary, the rolling quintessence scalar which couples to a hidden sector gauge field can lead to a time variation of the fine structure constant only if the hidden gauge field is massless\footnote{The bounds on dark radiation disfavour the presence of such massless gauge fields.}. Even in this case, this there is no tension with the 
bound for generic estimates of the loop factors.

\section{Discussion}
\label{sec:disc}

Geometric separation in the extra-dimensions provides a mechanism to have fields which interact with the visible sector with weaker-than-Planck suppressed couplings. We have examined the strength of such interactions in cases in which they arise due to kinetic mixing of scalars and gauge fields. In our explicit analysis of kinetic mixing of scalars we considered the prototypical settings to have geometric separation between a light modulus and the Standard Model: the modulus was 
taken to be a geometric blow-up modulus or a fibre modulus, while  the Standard Model was realised from D7-branes wrapping another blow-up mode or from branes at singularities. In all cases, we found that the bounds from fifth forces imposed interesting constraints. We would now like to make some general remarks.

From our computations in Section \ref{sec:kin}, it is easy to see that a non-zero $K_{\tau_q \tau_{\rm sm}}$ (where $\tau_q$ denotes a light scalar and $\tau_{\rm sm}$ a modulus that sets the value of the Standard Model couplings) entry in the \Kahler metric leads to
interactions between $\tau_q$ and the Standard Model sector. In cases where  this entry is vanishing, interactions will in general  be ``mediated" by the volume modulus. Since the wavefunction of the  volume modulus has support in all regions of the extra-dimensions, all moduli are expected to have direct interactions with it leading to non-zero  $K_{\tau_b \tau_{q}}$ and $K_{\tau_b \tau_{\rm sm}}$ entries in the kinetic matrix. Now, even if the $K_{\tau_q \tau_{\rm sm}}$ entry is non-zero, diagonalising the kinetic matrix will in general involve a basis change which leads to a mixing between $\tau_{\rm q}$ and $\tau_{\rm sm}$ (as can be seen from second-order perturbation theory in the off-diagonal entries of the \Kahler metric). An interesting exception to this is the case when $\tau_q$ is
exactly at the singular locus. At leading order, the \Kahler potential for the field is given by:
\begin{equation}
\label{singk}
   K \supset { \tau_q^2 \over \V }.
\end{equation}
Note that $K_{\tau_q \tau_b}$ vanishes at $\tau_q = 0$, thus interactions ``mediated by the volume" are absent. This
vanishing is similar in spirit to the mechanisms in \cite{Damour:1994ya, Damour:2002mi, Brax:2009kd} to avoid couplings between moduli and Standard Model fields. Interactions would be induced by terms in the \Kahler potential which are linear in $\tau_q$, such as:
\begin{equation}
\label{orbmix}
   K \supset {\tau_q \tau_{\rm sm}  \over \V^{p}} \ .
\end{equation}
It should be possible to determine the power of volume $(p)$ that appears in Eq. \eqref{orbmix} by performing calculations in the orbifold limit. If examples with high values of $p$ can be found (see \cite{Kobayashi:2011cw, Maharana:2011wx} for symmetry considerations that can lead to high $p$), they would provide interesting settings to evade bounds from fifth forces. Although, if the field is to be used to drive quintessence one would have to explain why the scalar is  exactly at or very close to the singular point today. More generally, there are various mechanisms  to realise sequestered sectors. Fields localised in warped throats interact weakly with degrees of freedom in the bulk of the compactification. In addition to suppression by powers of the volume\footnote{We note that the wavefuction of the volume modulus becomes non-uniform in the presence of warping, with lower support in warped throats \cite{Giddings:2005ff, Frey:2006wv}.}, the interactions are suppressed by the warp factor at the bottom of the throat. The
construction of \cite{Panda:2010uq} uses an axionic field in a $10^{-3} \, \rm {eV}$ warped throat to drive quintessence\footnote{Axions do not mediate long-range forces between macroscopic bodies, fifth forces are trivially satisfies for light axions, \cite{Panda:2010uq} used warping to lower the scale of the quintessence potential.}. As discussed in detail in \cite{Panda:2010uq}, there are many model building challenges that can arise in constructions with such long throats: the cosmological moduli problem, the danger of formation of black brane horizons and overproduction of dark radiation. These have to be addressed in detail for each model separately. Another way to evade the bounds from fifth forces is to construct models where  screening effects  \cite{Khoury:2010xi, Vainshtein:1972sx, Brax:2004qh, Brax:2004qh, ArmendarizPicon:2000dh, Hinterbichler:2010es, Martin:2008qp} are relevant. See \cite{Brax:2006np} for a discussion of possible embedding of the chameleon mechanism in supergravity\footnote{For general discussions of quintessence model building in supergravity see e.g. \cite{Brax:1999gp, Copeland:2000vh, Chiang:2018jdg, Kallosh:2002gf}.}.
Finally, we would like to mention that in arriving at the lower bounds on the volume we have assumed that there is no alignment between
the kinetic and mass matrices so that their effects precisely cancel when the interactions between the light modulus and the visible sector 
are computed (as described in detail in Section \ref{sec:kin}). It will be interesting to explore if it is possible to get such alignments in string
compactifications naturally.

Next, we would like to discuss another interesting feature revealed by  our study. In many of the examples in Section \ref{sec:kin}, we have found the strengths of the couplings of the geometrically separated modulus to the visible sector  fields to be non-universal, i.e. it couples to the gauge bosons and matter fermions with  different strengths. The implications of such couplings for violations of the equivalence principle have not been studied in detail in the literature. It is important to develop the detailed phenomenology of such models as in \cite{Kap}, starting from the RG running of the high scale Lagrangian. We hope to pursue this direction in the future.

Our results can also be used to quantify the fine-tuning necessary for stability against quantum corrections involving visible sector loops. A rough estimate of the effect of visible sector loops on the mass of the quintessence field can be obtained as in \cite{Adelberger:2003zx}, for a theory  where the couplings between the quintessence field and the visible sector fields are suppressed by the scale $\Lambda$, quantum corrections give
\begin{equation}
\label{eq:MassCorrection}
\delta m_q^2 \sim {\Lambda_{\rm UV}^4 \over \Lambda^2},
\end{equation}
where $m_q$ is the mass of the quintessence field and $\Lambda_{\rm UV}$ is the cut-off scale. For the example involving two blow-up moduli discussed in Section \ref{twoblow}, if the volume is taken to be $\V \sim 10^{12}$ (so that the fifth force bounds are evaded), Eq. \pref{eq:MassCorrection} yields $\delta m_q \sim 1 \ {\rm GeV}$, where we have taken supersymmetry to be broken at a high scale (i.e there are no cancellations amongst visible sector loops). This is forty orders of magnitude greater than the physical mass. Similar estimates can be performed for the other examples and scenarios with low scale supersymmetry\footnote{Supersymmetry together with weaker-than-Planck suppressed interactions can lead to unexpectedly light scalars in extra-dimensional theories \cite{uber}.}. For more accurate quantifications and an understanding of the functional fine-tuning involved, one can start from the couplings derived in  Section~\ref{sec:kin} and make use of the formalism developed in \cite{Garny:2006wc}.

\section{Conclusions}
\label{sec:conc}

Moduli fields play a central role in string phenomenology. For Planck suppressed interactions between  moduli and  the visible sector,  fifth force bounds  prevent them from being light and the bounds from time variation of fundamental constant prevent them from being cosmologically active. Geometric separation can lead to weaker-than-Planck suppressed couplings between moduli and the visible sector. We have examined the strength of the interactions between such geometrically separated moduli and the Standard Model sector induced by kinetic mixings. Our results should provide lower bounds on the strength of such interactions unless the mass and kinetic matrices are aligned so as to cancel each others effects or screening effects are relevant. If the modulus is taken to be massless (which is a good approximation if it is to drive quintessence) fifth force bounds lead to interesting lower bounds on the volume. In the context of quintessence, our results reiterate the importance of constructing models where all moduli are stabilised, so that all the couplings of the field driving  quintessence can be computed explicitly and the compatibility with fifth force bounds can be examined. More generally, the next generation of experiments plan to improve on the tests of the equivalence principle by two orders of magnitude (see e.g. \cite{Nobili:2018eym}) - the time is ripe to develop a detailed understanding of couplings between moduli fields and the visible sector fields in string theory models.

\section*{Acknowledgements}

We would like to thank  Michele Cicoli, Joe Conlon,  Shanta de Alwis, Fernando Quevedo and Ivonne Zavala for useful discussions and communications. The work of BSA was supported by the STFC Grant ST/L000326/1. This work was supported by a grant from the Simons Foundation (\# 488569, Bobby Acharya).

\end{document}